\documentclass{PoS}

\usepackage{epstopdf}
\newcommand{\snn}{\sqrt{s_{\rm NN}}}

\newcommand{\pt}{p_{\rm T}}

\newcommand{\raa}{R_{\rm AA}}

\newcommand{\raapi}{\raa^{\pi^{\pm}}}
\newcommand{\raaD}{R^{\rm D}_{\rm AA}}
\newcommand{\raaB}{R^{\rm B}_{\rm AA}}

\title{Overview on heavy flavour measurements in lead-lead collisions at the CERN-LHC}

\ShortTitle{Overview on heavy flavour measurements at the CERN-LHC}

\author{\speaker{Andr\'e Mischke} \\
       ERC-Research Group {\em QGP-ALICE}, \\ Utrecht University, 
       Princetonplein 5, 3584 CS Utrecht, the Netherlands \\
       E-mail: \email{a.mischke@uu.nl}}

\abstract{High energy collisions of heavy atomic nuclei allow to create and carefully study a high-density, colour-deconfined state of strongly-interacting matter. According to calculations from lattice Quantum-Chromodynamics, under the conditions of high energy density and temperature reached in such collisions, the phase transition to a quark-gluon plasma (QGP) is expected to occur, where the colour confinement of quarks and gluons into hadrons should vanish and chiral symmetry should be restored. Heavy-flavour particles, containing charm and beauty, are unique probes of the conditions of the medium formed in nucleus-nucleus collisions at high energy.
In this report recent measurements on open and hidden heavy-flavour production in lead-lead collisions at CERN's Large Hadron Collider are presented and discussed.}

\FullConference{X$^{th}$ Quark Confinement and the Hadron Spectrum, \\
		October 8-12, 2012 \\
		TUM Campus Garching, Munich, Germany}

\begin{document}

%
\section{Introduction}
\begin{figure}[b]
  \centering
  \includegraphics[width=0.8\textwidth]{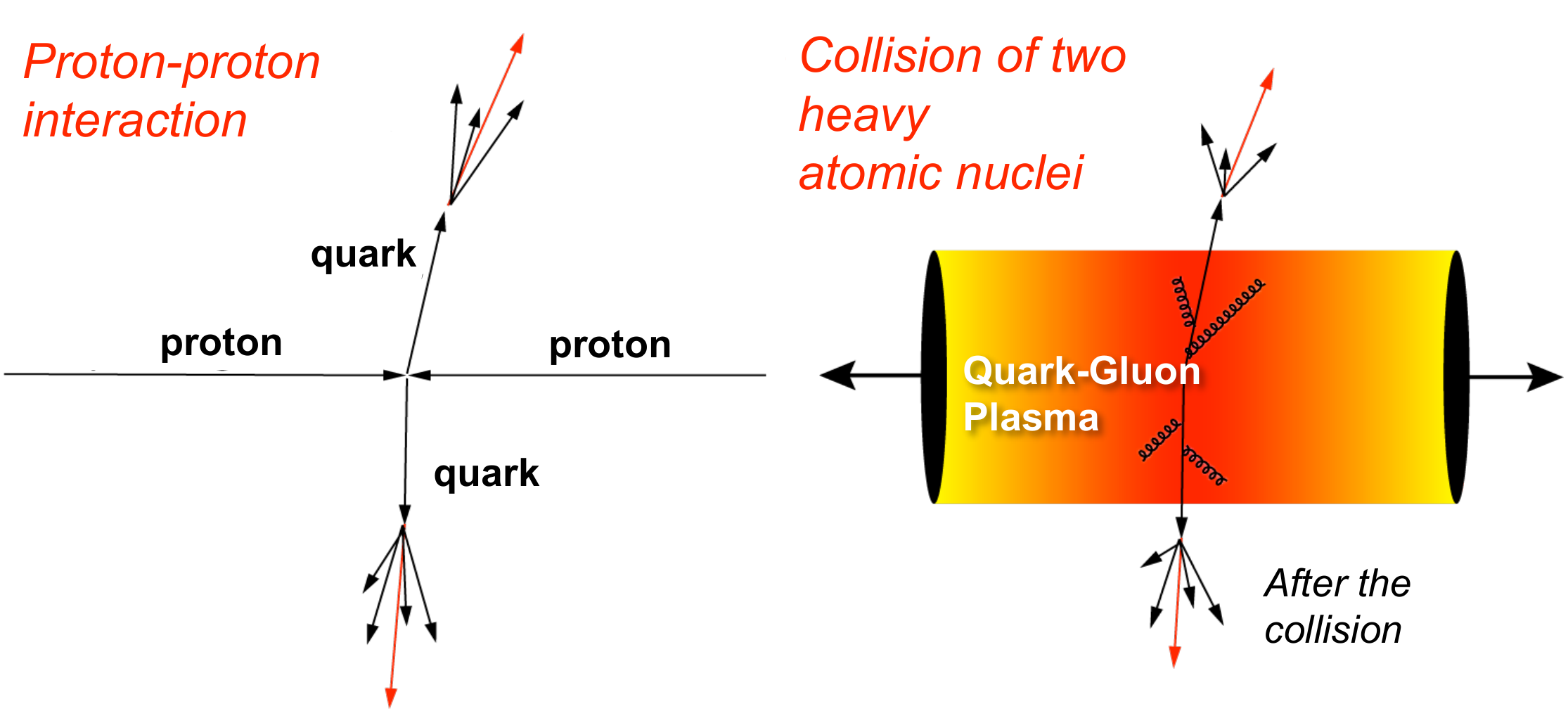}
  \vspace*{-0.4cm}
  \caption{Sketch of an elementary proton-proton interaction (left panel), which produces two back-to-back orientated sprays (jets) of collimated particles from the hard scattering of two quarks or gluons. In a collision of two heavy atomic nuclei (right panel) the elementary interaction takes place slightly before the creation of the QGP matter. The hard scattered quarks traverse through the QGP, interact strongly with it and loose energy through gluon radiation.}  
\label{Fig:1}
\end{figure}
Quantum Chromodynamics (QCD) is well established as the fundamental theory describing the most powerful force of nature, the strong interaction between quarks and gluons. These quarks and gluons are the building blocks of atomic nuclei and account for over 99\% of the visible mass of the universe. QCD successfully explains a wide range of phenomena, from the mass spectrum of hadrons to deep-inelastic scattering processes. One of the characteristic features of QCD is asymptotic freedom where the interaction becomes weak at large momentum transfer, in other words, when the colour charged quarks come close to each other. In contrast, the potential between the quarks increases steeply when they are separated. In fact, the binding force becomes so strong that, in our normal world, the quarks and gluons are permanently confined inside hadrons.

A fascinating and direct consequence of asymptotic freedom is that under the conditions of sufficiently high temperature or density the strongly interacting quarks and gluons are liberated from their hadronic confinement. This extraordinary new state of matter, where the basic degrees of freedom are released, is called the Quark-Gluon Plasma (QGP). In cosmology, it is believed that the early expanding universe consisted of such plasma approximately 10 microseconds after the Big Bang. This plasma subsequently underwent a phase transition where quarks and gluons became confined to form colourless hadrons, which represent observed particles today. Quark matter may still exist in the core of neutron stars, where the density could exceed the critical value of the phase transition. The properties of such matter are fundamental predictions of QCD and its study is one of the leading and most active fields in contemporary subatomic physics.
The aim of ultra-relativistic heavy-ion physics is to create QGP matter and carefully study its complex nature under controlled laboratory conditions (for recent reviews, see~\cite{Intro1, Intro2}). 

At the onset of the collision, the quark and gluon constituents of the incoming nuclei can undergo hard collisions. In elementary proton-proton interactions these hard-scattered quarks or gluons fragment in the vacuum into jets of collimated, high transverse momentum particles, whose properties can be measured in the detectors (see left panel of Figure~\ref{Fig:1}). In heavy ion collisions however, these very energetic quarks or gluons traverse through the formed QGP matter and get slowed down or absorbed, much like X-rays traversing a tissue sample (see right panel of Figure~\ref{Fig:1}). The properties of the QGP matter can be studied through the attenuation of these energetic jets~\cite{Intro3}. The concept is similar to the idea of computed tomography used in medical imaging.
Measurements at the RHIC particle accelerator have revealed a softening and broadening of jets inside the QGP matter~\cite{Intro1}. The observed jet attenuation is evidence of the extreme energy loss of quarks or gluons traversing a large density of colour charges (so-called jet-quenching)~\cite{Intro3} and reflects the extreme opacity of the QGP. First attempts have been made to determine the stopping power of the QGP matter.

In November 2010, the Large Hadron Collider (LHC) started operation with heavy ion beams, colliding lead nuclei at a centre-of-mass energy of 2.76 TeV per nucleon-nucleon pair. This opened a new era in ultra-relativistic heavy ion physics at energies exceeding previous accelerators by more than an order of magnitude. The initial energy density in the collision zone is about a factor of 3 higher than at the RHIC facility. The higher energy density allows thermal equilibrium to be reached more quickly and to create a relatively long-lived QGP phase. Therefore, most of the in-medium effects should be enhanced, and this has already been observed at the LHC. Precise studies of the jet modification in the QGP matter were performed and important aspects of the jet-quenching theory have been tested over a much broader dynamic range than before~\cite{Intro4}. \\

Of particular interest is the dependence of the parton energy loss on colour charge and quark mass~\cite{Intro5}, which allows to gain more information about the dynamical properties of the QGP matter. For this purpose heavy quarks (charm and beauty) are the ideal probes. Charm quarks are about 250 times heavier than the light up and down quarks that dominate the QGP matter and their mass is not affected by chiral symmetry breaking~\cite{Intro6}. Beauty quarks are even three to four times heavier than charm quarks (but much less abundant). These large masses mean that charm and beauty quarks have much higher penetrating power than light quarks.

Due to their large mass, heavy quarks are produced predominantly in the (hottest) initial phase of the collision via gluon fusion processes~\cite{hf1} and therefore probe the complete space-time evolution of the QGP matter. As RHIC measurements have shown~\cite{hf2, hf3}, heavy-quark production by initial state gluon fusion dominates in heavy ion collisions where many (in part overlapping) nucleon-nucleon collisions occur. Thermal processes later in the collision might contribute to heavy-quark production at low transverse momentum~\cite{hf4}.


The particle production yield in heavy ion collision has contributions from initial- and final-state effects. They can be quantified using the nuclear modification factor $\raa$, where the particle yield in heavy ion collisions is divided by the yield in proton-proton reactions scaled by the number of binary collisions. $\raa$ = 1 would indicate that nucleus-nucleus collisions can be considered as an incoherent superposition of nucleon-nucleon interactions. Initial-state effects, such as Cronin enhancement, nuclear shadowing and gluon saturation~\cite{IniStatEff1, IniStatEff2, IniStatEff3, IniStatEff4}, give an $\raa$ different from 1. Final state effects, such as radiative and collisionial energy loss in the QGP matter, result in an $\raa$ smaller than unity~\cite{Eloss1}.
By comparing the nuclear modification factor of charged pions $(\raapi$), mostly originating from gluon fragmentation at this collision energy, with that of hadrons with charm $\raaD$ and beauty $\raaB$ the dependence of the energy loss on the parton nature (quark/gluon) and mass can be investigated.\\

A description of the Large Hadron Collider and the setup and performance of the ALICE and CMS experiments can be found in~\cite{lhc-det}. The details of the analysis for the measurement of D mesons and heavy-flavour decay electrons and muons are discussed in~\cite{andrea-hep, ralf}. A detailed discussion of the Quarkonia measurements can be found in~\cite{antoine, mironov}.

%
\section{Open heavy flavour production in pp and Pb--Pb collisions}
\begin{figure}[t]
  \centering
  \includegraphics[width=0.53\textwidth]{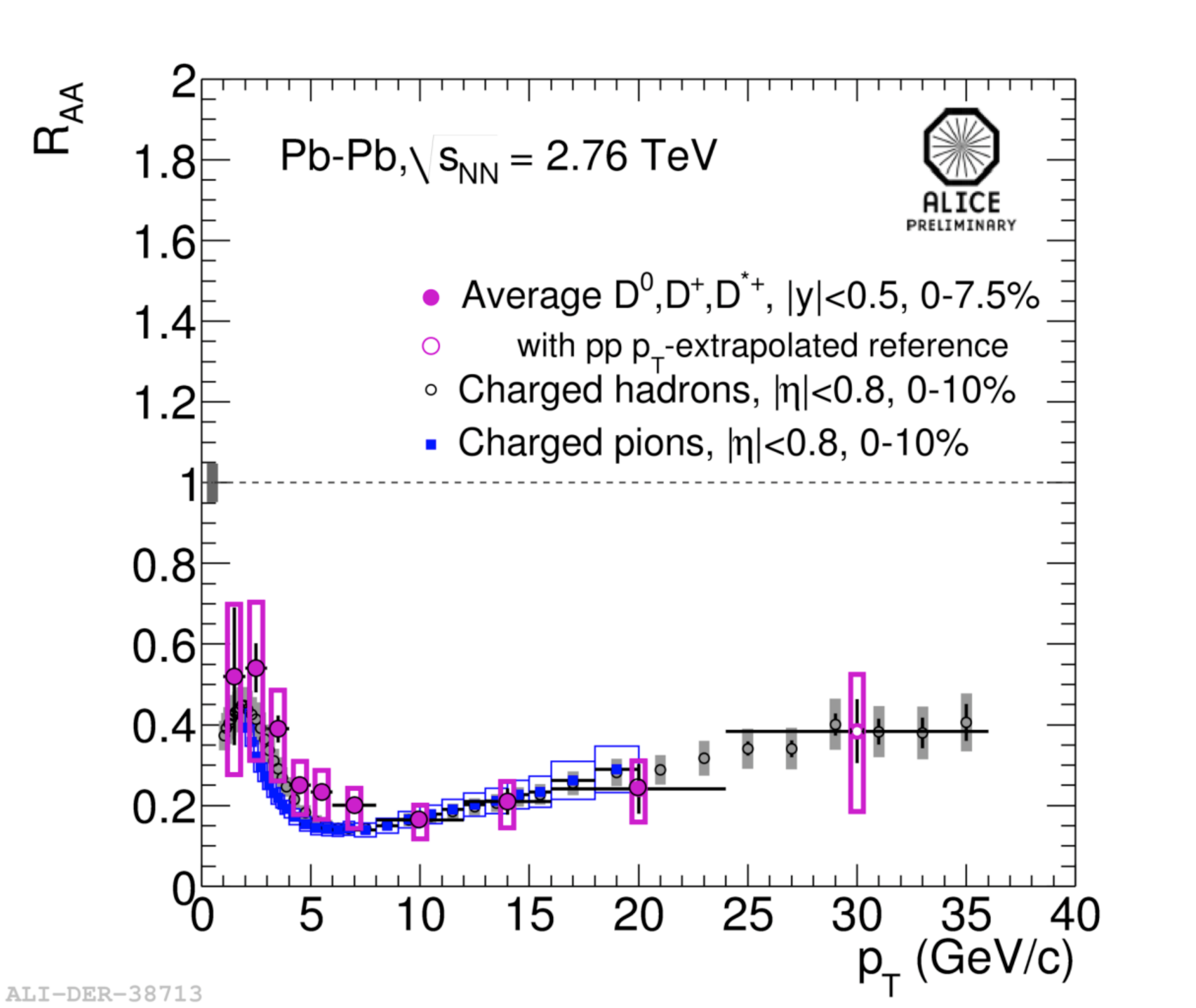}  
  \includegraphics[width=0.45\textwidth]{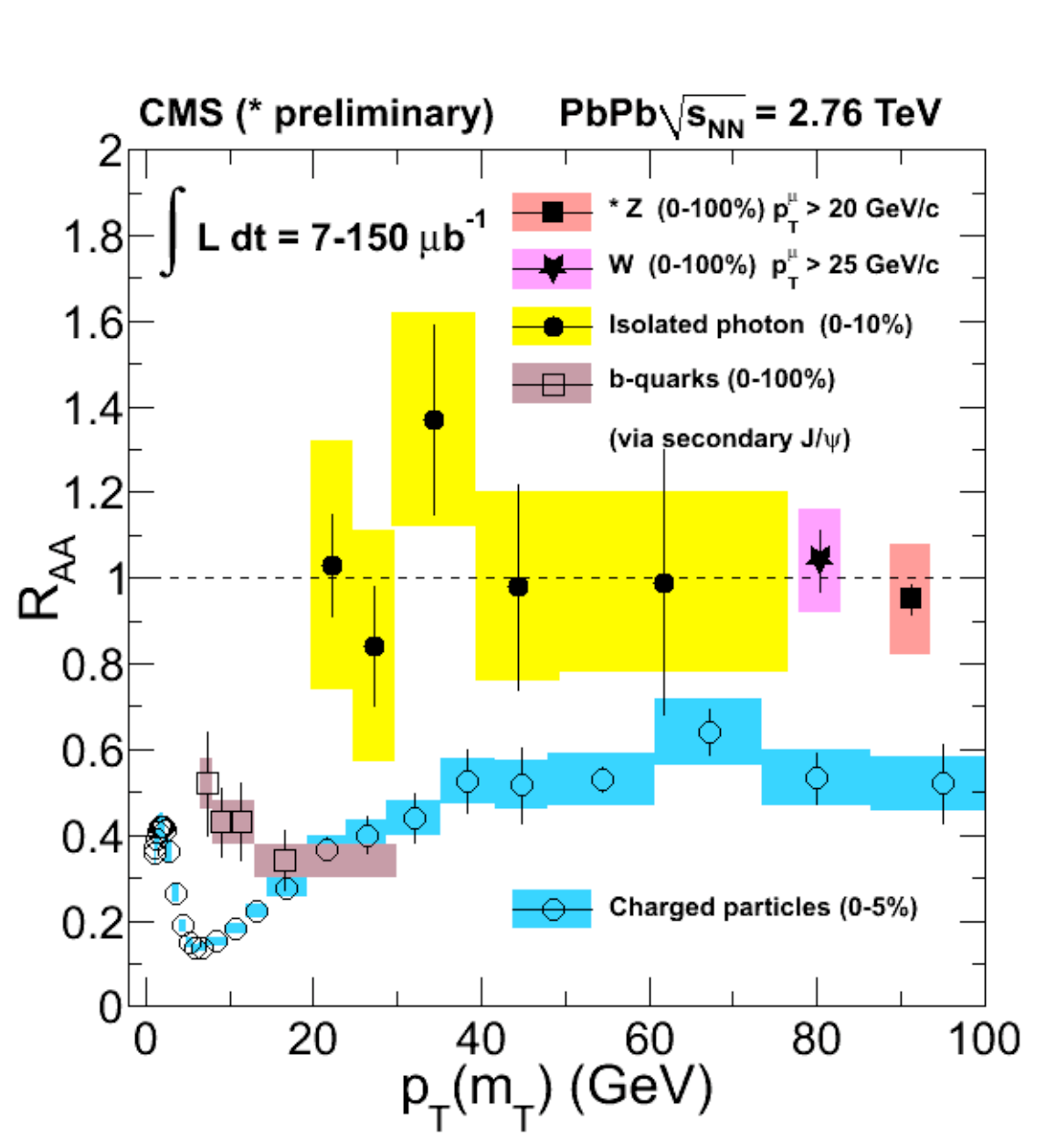}
  \caption{Left: Transverse momentum dependence of the average $\raa$ of prompt D mesons in the 0--7.5\% centrality class compared to the nuclear modification factors of charged hadrons and pions in lead-lead collisions at $\snn$~= 2.76 TeV. Right: $\raa$ of J/$\psi$ from B decays in minimum bias Pb--Pb collisions compared to the $\raa$ of charged particles (5\% most central collisions) and gauge bosons ($\gamma$, Z and W).}  
\label{Fig:2}
\end{figure}
The scattered heavy quarks loose energy when traversing through the QGP by medium-induced gluon radiation and collisions with the light quarks in the QGP. This interaction provides more insight on transport properties in the QGP and thus in the energy loss mechanisms.
Theoretical models based on perturbative QCD predicted that heavy quarks should experience smaller energy loss than light quarks due to the suppression of gluon radiation at small angles (so-called "dead-cone effect")~\cite{Eloss2, Eloss3, Eloss4}. Since this angle is mass dependent one expect less energy loss for charm hadrons at low transverse momentum compared to light quark hadrons (e.g. pions). Moreover, beauty is expected to be less suppressed than charm. Thus, the nuclear modification factors of hadronic yields should show a mass ordering pattern $\raapi<\raaD<\raaB$.

Surprisingly, measurements in head-on heavy-ion collisions from the ALICE collaboration and the RHIC experiments indicate that the yield of leptons from heavy-quark decays and prompt open charmed mesons are suppressed at the same level as observed for light-quark hadrons, which was not expected due to the dead-cone and colour-charge effects (cf. Figure~\ref{Fig:2}, left). Energy loss models currently describe the observed suppression at high transverse momentum reasonably well whereas the description at low transverse momentum ($\leq$ 2 GeV/$c$) is more challenging.

Measurements from the CMS collaboration based on displaced J/$\psi$ production and beauty-tag jets provide first indications that beauty has indeed smaller energy loss in the QGP than light quark hadrons (cf. Figure~\ref{Fig:2}, right). The colorless gauge bosons (isolated photons, Z and W) do not interact with the QGP matter and therefore have an $\raa$ of 1. These measurements also show that their production yield scales with the number of binary collisions.

\begin{figure}[t]
  \centering
  \includegraphics[width=0.57\textwidth]{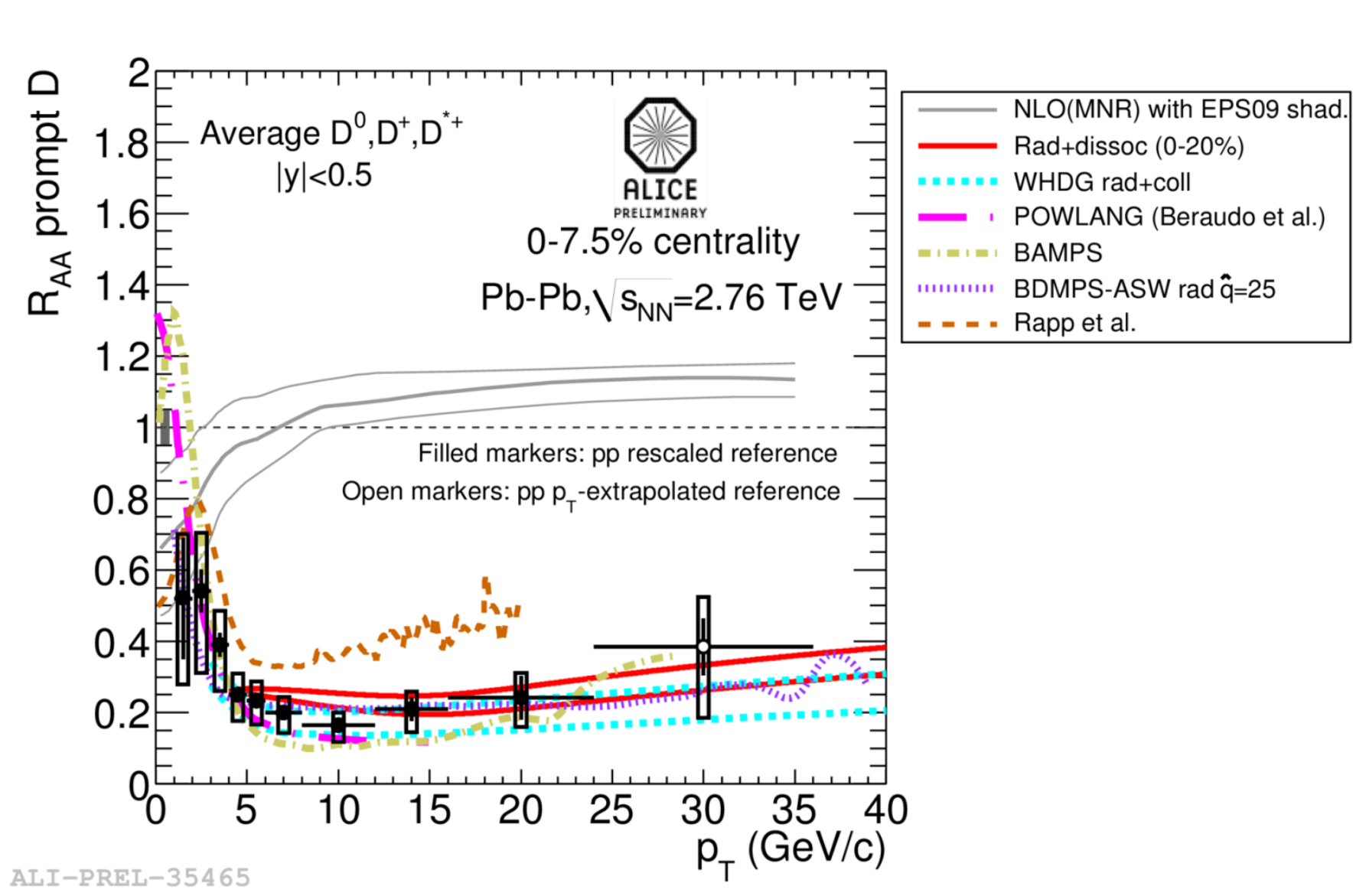}
  \includegraphics[width=0.423\textwidth]{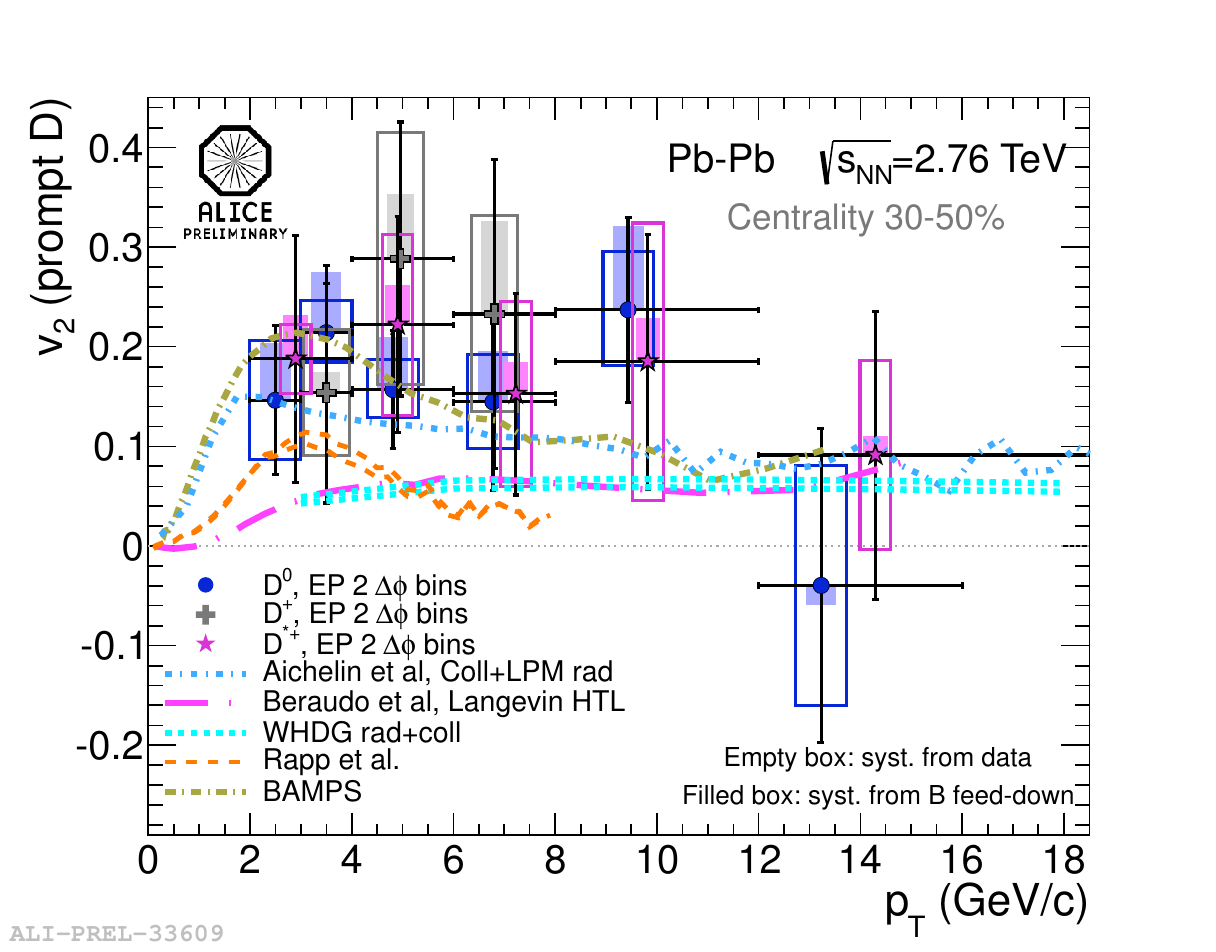}
  \vspace*{-0.3cm}
  \caption{Average $\raa$ and elliptic flow parameter $v_2$ of prompt D mesons at mid-rapidity in 7.5\% most central lead-lead collisions at $\snn$~= 2.76 TeV compared to next-to-leading-order perturbative QCD calculations with nuclear shadowing~\cite{D-NLO} and different parton energy-loss models~\cite{DmesonPbPb}.}  
\label{Fig:3}
\end{figure}

Furthermore, measurements of the momentum distribution of emitted particles and comparison with hydrodynamic model calculations have shown that the outward steaming particles move collectively, with the patterns arising from variations of pressure gradients early after the collision. This phenomenon is called azimuthal anisotropy or elliptic flow and is analogous to the properties of fluid motion. 
The study of the elliptic flow (or azimuthal anisotropy) of heavy-quark particles is particularly interesting as it provides information on the degree of thermalisation (interactions) of heavy quarks in the QGP. Sensitive ALICE measurements of the azimuthal anisotropy of electrons from heavy-flavour decays and prompt open charmed mesons in peripheral Pb--Pb collisions indicate a sizable flow of heavy quarks~\cite{andrea-hep}.
A simultaneous description of the nuclear modification factor and elliptic flow heavy-quark particles is challenging for the currently available theoretical model calculations (cf. Figure~\ref{Fig:3}).

%
\section{Hidden heavy flavour production in pp and Pb--Pb collisions}
\begin{figure}[t]
  \centering
  \includegraphics[width=0.45\textwidth]{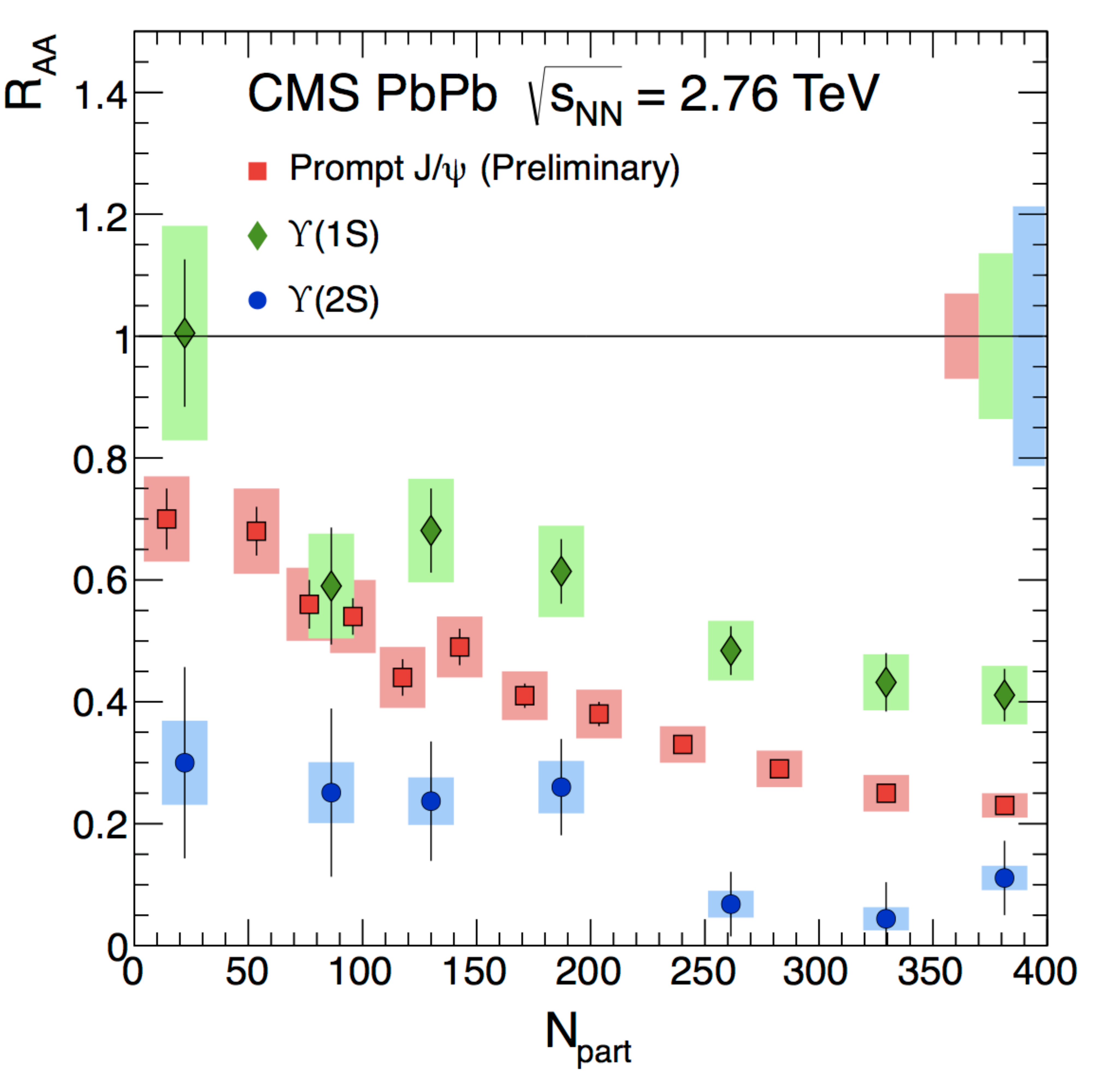}  
  \includegraphics[width=0.5\textwidth]{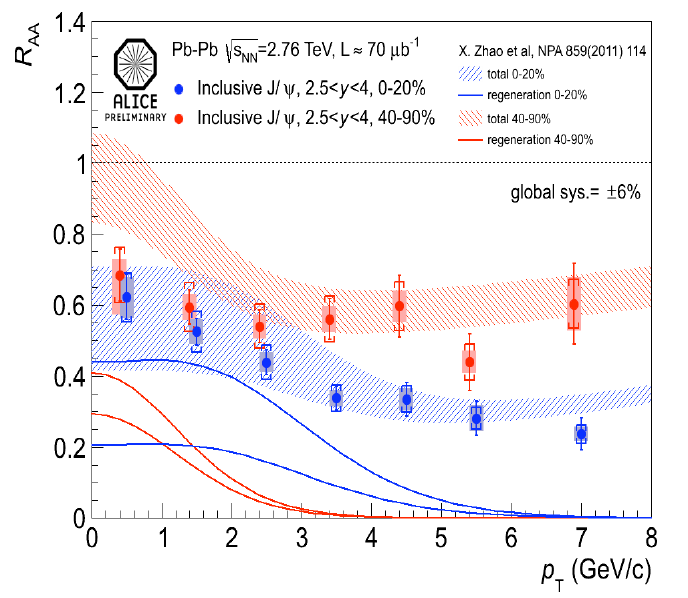}
  \vspace*{-0.3cm}
  \caption{Left: Centrality dependence of the nuclear modification factors for the prompt J/$\psi$, $\Upsilon$(1S) and $\Upsilon$(2S) states, measured by the CMS experiment in lead-lead collisions at $\snn$~= 2.76 TeV.
  Right: Transverse momentum dependence of the nuclear modification factor of inclusive J/$\psi$ for the centrality classes 0-20\% and 40-90\% measured with the ALICE experiment (symbols). The hashed bands show the overall production obtained from transport model calculation, while the open ones indicate the component from QGP regeneration.}  
\label{Fig:4}
\end{figure}

The dissociation of quarkonium states (hidden charm and beauty) due to colour screening in the QGP is one of the "classic signatures" of deconfinement~\cite{satz-1986}. A sequential suppression of the quarkonium states, such as $\Upsilon$(1S), $\Upsilon$(2S) and $\Upsilon$(3S), depends on their binding energy and the temperature of the surrounding medium, thus providing a so-called "QCD thermometer"~\cite{digal-2001}.
However, it has been shown that yield enhancement via subsequent regeneration in QGP or at chemical freeze-out of quarkonium states due to the large heavy-quark multiplicity might play an important role at LHC energies~\cite{hhf1, hhf2, hhf3, hhf4}. Furthermore feed-down from higher quarkonium states has to be considered~\cite{georg}.

The nuclear modification factor $\raa$ for different quarkonium states in lead-lead collisions at $\snn$~= 2.76 TeV indicates indeed a sequential melting of these states. Figure~\ref{Fig:4} (left panel) depicts the centrality dependence of the $\raa$ of the prompt J/$\psi$, $\Upsilon$(1S) and $\Upsilon$(2S) states, measured by the CMS collaboration.
The data support the hypothesis of increased suppression of less strongly bound states: the $\Upsilon$(1S) is the least suppressed and the $\Upsilon$(2S) is the most suppressed of the three states. 
The $\Upsilon$(1S) and $\Upsilon$(2S) suppressions are observed to increase with collision centrality. The observed $\Upsilon$(nS) yields contain contributions from heavier bottomonium states decays. As a result, the measured suppression is affected by the dissociation of these states. 
These results indicate that the directly produced $\Upsilon$(1S) state is not significantly suppressed, however quantitative conclusions will require precise estimations of the feed-down contribution matching the phase space of the suppression measurement. 

The ALICE experiment measured inclusive J/$\psi$ production in central Pb--Pb collisions at $\snn$~= 2.76 TeV down to $\pt$~= 0 (cf. Figure~\ref{Fig:4}, right). Less suppression was found at LHC both at forward rapidity and at mid- rapidity. Moreover, an enhanced production at low $\pt$ compared to high $\pt$ was observed, a trend which is, again, different from the one seen at RHIC.
These observations support the picture of J/$\psi$ melting in the QGP phase, followed by subsequent regeneration in the QGP or at chemical freeze-out. Both the statistical hadronization~\cite{jpsi1} and transport models~\cite{jpsi2, jpsi3} assume thermalization of charm quarks in the QGP and reproduce the data. Those transport model calculations predict that about half of the low-$\pt$ J/$\psi$ yield is produced by the recombination of charm quarks in QGP, while the rest is due to primordial production. Within the statistical hadronization model~\cite{jpsi1}, the charmonium states probe the phase boundary between the QGP and the hadron phase. This extends with a heavy quark the family of quarks employed for the determination of the hadronization temperature (via the conjectured connection to the chemical freeze-out temperature extracted from fits of statistical model calculations to hadron abundances).

%
\section{Summary and outlook}
Heavy quarks (charm and beauty) are sensitive penetrating probes to study the dynamical properties of the quark gluon plasma (QGP), created in ultra-relativistic heavy-ion collisions. Due to their large mass (larger than about 1.3~GeV/$c^2$), they are predominantly produced in the early stage of the collision by gluon-fusion processes, so that they provide information about the hottest initial phase.

Open and hidden heavy-flavor production has been studied in pp and Pb--Pb collisions at the CERN-LHC.
The nuclear modification factor of prompt D$^0$, D$^+$, D$^{*+}$ and D$_{\rm s}^{\rm +}$ mesons in the most central Pb--Pb collisions are suppressed to the same level as observed for light-quark hadrons. 
A sequential suppression of the individual $\Upsilon$ states in Pb--Pb collisions at the LHC with respect to their yields in pp data has been observed.
The inclusive J/$\psi$ production is less suppressed at low transverse momentum, which was not observed at RHIC collision energy. Comparison with transport and statistical model calculations suggest that a sizeable regeneration component is needed to describe the data at low transverse momentum.

The observed signals for the QGP are expected to be even stronger in Pb--Pb collisions at $\snn$~= 5.5 TeV and allow the properties of the QGP to be further characterized. 
Proton-lead data are urgently needed to measure the contribution from the effects of cold nuclear matter, such as nuclear shadowing and Cronin enhancement. 
The experimental teams at the LHC and at RHIC are working on upgrades of the inner tracking systems of their detectors, aiming for an improved resolution in impact parameter, which will allow to measure charm baryons and will improve the precision on the current open charm and open beauty measurements.

\section*{Acknowledgments}
I would like to thank the organisers for inviting me and the stimulating discussions at this conference. 
Moreover, I thank the ALICE and CMS Collaborations for the data and the LHC accelerator team. \\
The European Research Council has provided financial support under the European Community's Seventh Framework Programme (FP7/2007-2013) / ERC grant agreement no 210223.
This work was also supported by a Vidi grant from the Netherlands Organisation for Scientific Research (project number 680-47-232) and a Projectruimte grant from the Dutch Foundation for Fundamental Research (project number 10PR2884).


\end{document}